# End-to-End Deep Diagnosis of X-ray Images

Kudaibergen Urinbayev, Yerassyl Orazbek, Yernur Nurambek, Almas Mirzakhmetov and
Huseyin Atakan Varol, *Senior Member, IEEE*

*Abstract*— In this work, we present an end-to-end deep learning framework for X-ray image diagnosis. As the first step, our system determines whether a submitted image is an X-ray or not. After it classifies the type of the X-ray, it runs the dedicated abnormality classification network. In this work, we only focus on the chest X-rays for abnormality classification. However, the system can be extended to other X-ray types easily. Our deep learning classifiers are based on DenseNet-121 architecture. The test set accuracy obtained for 'X-ray or Not', 'X-ray Type Classification', and 'Chest Abnormality Classification' tasks are 0.987, 0.976, and 0.947, respectively, resulting into an end-to-end accuracy of 0.91. For achieving better results than the state-of-the-art in the 'Chest Abnormality Classification', we utilize the new RAdam optimizer. We also use Gradient-weighted Class Activation Mapping for visual explanation of the results. Our results show the feasibility of a generalized online projectional radiography diagnosis system.

Keywords: Chest X-ray images, computer-aided diagnosis, digital radiography, deep learning, neural networks, explanatory visualization

## I. Introduction

In recent years, deep learning (DL) has demonstrated outstanding performance in a variety of tasks ranging from recognizing handwritten digits to defeating the world champions in multiplayer online games [1]. One of the areas that benefited most from the advances in DL is healthcare. For instance, deep natural language processing (NLP) was utilized for the automated analysis of electronic health records (EHRs). Reinforcement learning was leveraged for improved robotic-assisted surgery (RAS). Meanwhile, image processing and understanding using DL led to huge strides in radiology [2].

Even though DL in medical imaging has many applications such as image quality improvement [3], image registration [4], and organ segmentation [5] [6], the most prevalent use case is the computer-aided diagnosis (CAD). Before DL, machine learning for CAD relied on handcrafted feature engineering [7]. Moreover, the computational cost was a barrier to implementation. Currently, DL can extract domain-dependent features automatically. Also, advances in parallel computing, specifically graphical processing units (GPUs), increased the computational resources substantially.

This work was supported by the Institute of Smart Systems and Artificial Intelligence (ISSAI).

K. Urinbayev, Y. Orazbek, Y. Nurambek and H. A. Varol are with the Dept. of Robotics and Mechatronics, Nazarbayev University, 53 Kabanbay Batyr Ave, Z05H0P9 Nur-Sultan City, Kazakhstan. Email: {kudaibergen.urinbayev,yerassyl.orazbek, yernur.nurambek,ahvarol}@nu.edu.kz

Corresponding author: Huseyin Atakan Varol.

The CAD systems could be used for different medical imaging modalities such as magnetic resonance imaging (MRI), computed tomography (CT), positron emission tomography (PET), ultrasound, and projectional radiography. Projectional radiography, commonly referred to as X-Ray, is a two-dimensional medical image modality that is widespread in clinical practice as an affordable, noninvasive, and instant screening examination. It is used for diagnosing pulmonary diseases, broken bones, and other abnormalities. The common types of X-ray images are abdominal, barium, bone, chest, dental, extremity, hand, joint, lumbosacral spine, neck, pelvis, sinus, skull, thoracic spine, upper gastrointestinal and small bowel series [8].

Digital radiography captures the X-rays by imaging sensors in a digital format instead of a film in the traditional radiography. Picture archiving and communication systems (PACS) and EHRs in healthcare facilities enable secure storage and fast access to medical imaging data. Thanks to these, it is now possible to effortlessly get a digital copy of a radiography image for a second opinion. However, due to the high cost of the medical services and the shortage of qualified radiologists [9], it is still a challenge to get a confirmation of the results. Indeed, it might even be hard to get a primary diagnosis for those who cannot afford or access to medical services.

"Good Health and Well-being", and "Reduced Inequality" are two of the United Nations Sustainable Development Goals (SDGs) [10] that intend to provide universal health coverage and decrease poverty. Artificial intelligence can be utilized to achieve these goals. Taking into account the penetration rate of mobile networks, CAD can become a free online service accessible by all who have a smartphone with an Internet connection.

In this work, we present a generalized X-ray deep diagnostic pipeline (see Fig. I). This end-to-end framework can be further developed into an online CAD system. Such a tool could serve as a decision support system for physicians or urge patients with radiographs to seek professional medical support. Even though web-delivered diagnostic tools were presented in the literature, they are based on a single X-ray type, e.g. chest X-rays [11]. Our framework leverages the large-scale labeled X-ray datasets [12]–[16] which recently became available.

Our diagnosis framework in Fig. I consists of four main modules: The first module, "X-ray or not X-ray", differentiates radiography images from other types of images. The second module determines the X-ray type. The third module is an individual DL model for each X-ray type which identifies the abnormalities in each image. In this work, we present the chest X-ray module. The last module outputs the visually interpretable heatmap of the detected anomaly. The remainder of the paper is organized as follows: In Section II, we describe our data sources for different X-ray types. Our DL network architectures and their training procedures are explained in Section III. After presenting and discussing the DL results in Section IV, conclusions are drawn in Section V.

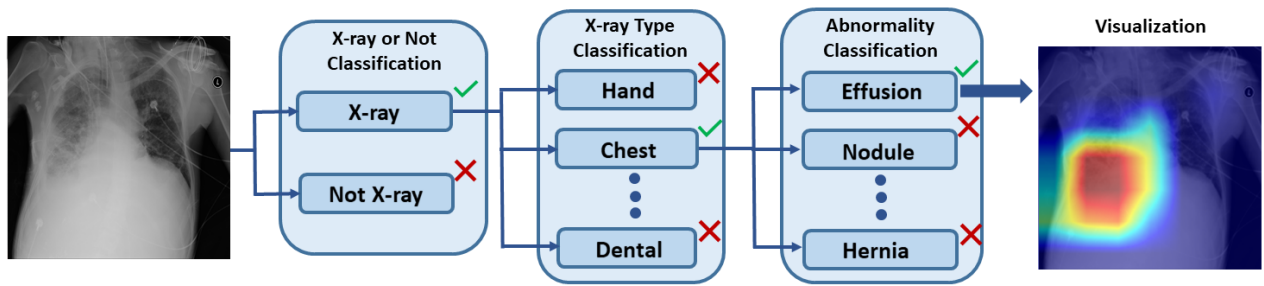

Fig. 1. Our DL pipeline for generalized X-ray diagnosis with explanatory visualization.

## II. DATA SOURCES

In order to train our DL networks, we used publicly available datasets.

### A. Chest-Xray14 dataset [12]

The dataset was used to train the models in all of the three modules and contains 112,120 weakly labeled frontal-view radiography images with 14 classes of diseases: Atelectasis, Cardiomegaly, Effusion, Infiltration, Mass, Nodule, Pneumonia, Pneumothorax, Consolidation, Edema, Emphysema, Fibrosis, Pleural Thickening, and Hernia. It should be noted that due to the NLP-based labeling of the images, this dataset contains errors in labeling. However, it is the largest dataset available for chest X-ray images.

### B. MURA (MUsculoskeletal RAdiographs) Dataset [13]

MURA dataset consists of 40,561 manually labeled images from 14,863 studies. It contains finger, wrist, elbow, forearm, hand, humerus, and shoulder radiographs. We used this dataset for training the 'X-ray or Not' and 'X-ray Type Classification' modules.

### C. LERA (Lower Extremity RAdiographs) Dataset [14]

LERA dataset was also used for training the 'X-ray or Not' and 'X-ray Type Classification' modules. The dataset contains 1,299 radiography images of the foot, knee, ankle, and hip collected from 180 studies.

### D. AASCE (Accurate Automated Spinal Curvature Estimation) Challenge Dataset [15]

The dataset released by SpineWeb contains 481 spinal anterior-posterior X-ray images, all of which present symptoms of scoliosis to some extent. The dataset of spinal X-ray images was utilized again in the training of 'X-ray or Not' and 'X-ray Type Classification' modules.

### E. Panoramic Teeth X-ray Dataset [16]

This dataset consists of 1,500 teeth X-ray images collected during extra-oral radiographic examinations. It was used to train the models in the first two modules.

### F. ImageNet [17]

This dataset was used to train the model 'X-ray or Not'. ImageNet is a comprehensive and diverse dataset of images with more than 14,197,122 images given in 21,841 synsets (i.e. sets of cognitive synonyms).

## III. METHODS

The first three modules ('X-ray or Not', 'X-Ray Type Classification', and abnormality classification) were implemented as classification models. These DL models used the ChexNet architecture [18] which is a customized variant of DenseNet-121 [19] for chest X-ray disease classification. The models were trained using Catalyst [20], a PyTorch [21] framework designed for reproducible results in research. NVIDIA DGX-1 server was used for model training.

### A. Module I: "X-ray or Not' Classification

The first module, which classifies X-ray images from other images, utilized all datasets mentioned in Section II to form the X-ray class. Only 10 percent of the chest X-ray images were used to prevent bias in the dataset resulting in over 50,000 samples for this class. As the 'Not X-ray' class, the same number of regular images were randomly chosen from the ImageNet dataset. The dataset was split to training, validation, testing sets with a ratio of 70%, 20%, and 10%, respectively. All images were converted to 8-bit grayscale and rescaled to 224x224 pixels. The training was carried out for 10 epochs using Adam optimizer (0.001 learning rate, betas (0.9, 0.999), and weight decay of 1e-5). Moreover, ReduceLROnPlateau, a dynamic learning rate scheduler, was utilized (Mode was set to "min", factor to 0.1, and patience to 3).

### B. Module II: X-ray Type Classification

This module was designed to classify different types of X-rays as listed in Table I. 14 X-ray classes were collected from five medical radiography datasets described in Section II. The data was also divided into 70% training, 20% validation, and 10% testing sets. The same model and training parameters were applied to the model as in Module I.

## C. Module III: Abnormality Classification

To train DenseNet-121 a new optimizer called RAdam [22] was used. It shows almost the same results with different initial learning rates, thus, the number of hyperparameters is reduced. It is achieved by implementing automated variance reduction and thus removes the need for manual tuning. Besides, developers of RAdam claim that it is more robust compared to Adam, SGD and RMSprop optimizers and less likely to converge into poor local minima. Binary Cross-Entropy Loss was used as a criterion. Transfer learning from ImageNet weights was performed.

The training was carried out for 15 epochs, with RAdam optimizer of 1e-4 learning rate and weight decay of 3e-4. The same ReduceLROnPlateau as in Module I & II was introduced.

The results of the abnormality recognition are explained visually using Gradient-weighted Class Activation Mapping (Grad-CAM) [23]. Grad-CAM helps to interpret the results of abnormality classification by creating class-discriminative visualizations. For a particular class of abnormality, Grad-CAM attributes importance to the neurons in the last convolutional layer of CNN by using the gradient information. Then, Grad-CAM combines the forward activation maps using weights and this operation is followed by a ReLU, resulting in a coarse heatmap. This heatmap highlights the features of the image that are important for predicting a particular abnormality.

## IV. RESULTS AND DISCUSSION

After the training of Densenet-121 for 'X-ray or Not' classification task, the accuracy for the testing set was obtained as 0.987. The corresponding confusion matrix is provided in Table II. Figure 2 shows examples of the misclassified X-ray and regular images.

Figure 3 demonstrates the confusion matrix of the X-ray type classification module. The accuracy for the testing set is 0.976. Most misclassifications are related to the musculoskeletal radiographs of similar body parts. For instance, 23 forearm X-rays are classified as elbow X-ray images.

The receiver operating characteristic (ROC) curves and aera under curve (AUC) scores in Fig. 4 show that different medical conditions can be identified with high accuracy.

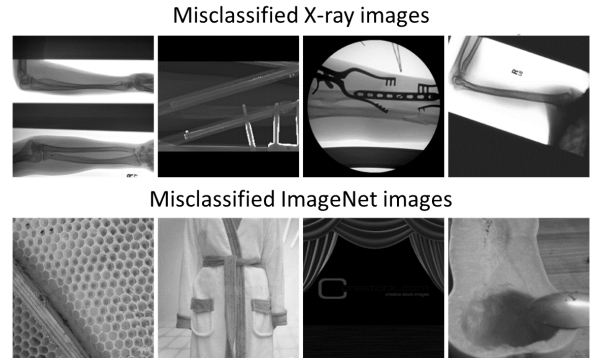

Fig. 2. Misclassified test samples from the 'X-ray or Not' classification task.

Fig. 3. Confusion matrix for X-ray type classification task.

TABLE I
TYPES AND NUMBER OF SAMPLES OF X-RAY IMAGES

| X-ray type | Spine | Elbow | Finger | Forearm | Hand |
|---|---|---|---|---|---|
| # of samples | 609 | 5396 | 5567 | 2126 | 6003 |
| X-ray type | Wrist | Knee | Foot | Ankle | Hip |
| # of samples | 10411 | 534 | 348 | 322 | 94 |
| X-ray type | Humerus | Shoulder | Dental | Chest | **Total** |
| # of samples | 1560 | 9054 | 1500 | 11001 | **54525** |

TABLE II
CONFUSION MATRIX OF THE X-RAY OR NOT CLASSIFICATION

|  | Predicted X-ray | Predicted ImageNet |
|---|---|---|
| **Actual X-ray** | 4712 | 5 |
| **Actual ImageNet** | 72 | 4928 |

TABLE III
AUC SCORES OF CHESTX-RAY14, CHEXNET, AND OUR DL MODEL

|  | ChestX-ray14 | CheXNet | Our Results |
|---|---|---|---|
| Atelectasis | 0.71 | 0.80 | 0.81 |
| Cardiomegaly | 0.80 | 0.92 | 0.91 |
| Effusion | 0.78 | 0.86 | 0.87 |
| Infiltration | 0.60 | 0.73 | 0.72 |
| Mass | 0.70 | 0.86 | 0.85 |
| Nodule | 0.67 | 0.78 | 0.78 |
| Pneumonia | 0.63 | 0.76 | 0.74 |
| Pneumothorax | 0.80 | 0.88 | 0.90 |
| Consolidation | 0.70 | 0.79 | 0.79 |
| Edema | 0.83 | 0.88 | 0.91 |
| Emphysema | 0.81 | 0.93 | 0.92 |
| Fibrosis | 0.76 | 0.80 | 0.81 |
| Pleural Thickening | 0.70 | 0.80 | 0.79 |
| Hernia | 0.76 | 0.91 | 0.99 |

However, the chest dataset is weakly labeled. Therefore, large scale X-ray datasets labeled by radiologists are needed to increase the reliability of the results. The accuracy of our model for abnormality identification is compared to the ChestX-ray14 and CheXNet models in Table III. Figure 5 demonstrates the explanatory visualization using Grad-Cam. For the X-ray with cardiomegaly label, the region around the enlarged heart is highlighted.

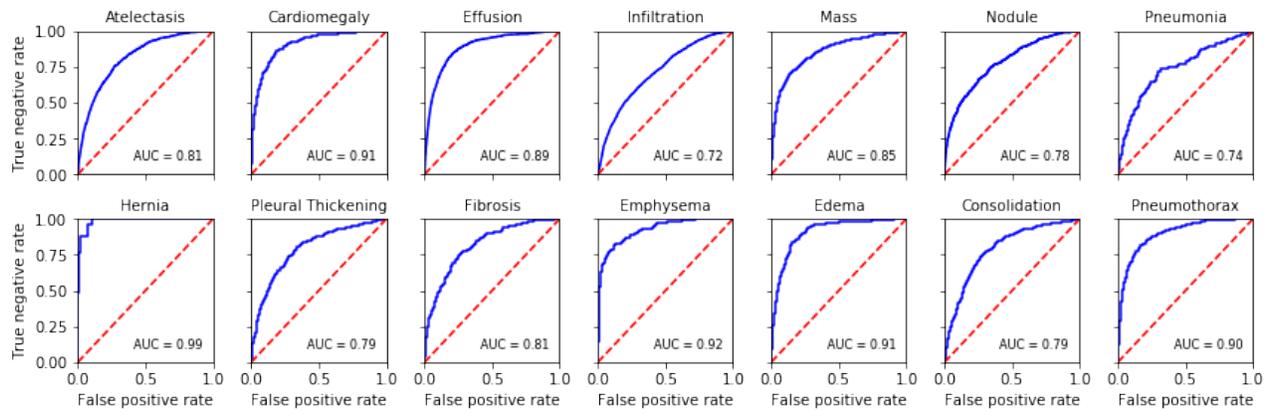

Fig. 4. ROC and AUC of 14 conditions on Chest-Xray14 testing set.

## V. Conclusion

The end-to-end accuracy of 91% for X-ray images show that data-driven CAD has a significant potential to assist physicians as a decision support tool. It can also allow individuals living in developing countries to get preliminary results for their projectional radiography examinations.

We intend to extend our work with diagnostic modules for other types of X-ray images. We also plan to develop our framework to an online and free tool accessible via personal computers and smartphones. The models described in this work ara avaliable for download at our website (https://issai.nu.edu.kz/xray).

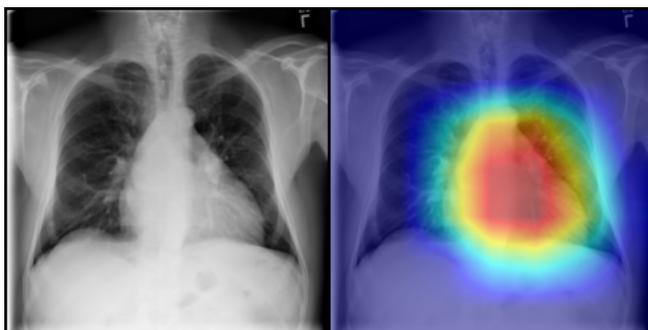

Fig. 5. Original X-ray image with cardiomegaly (left) and its explanatory visualization using Grad-Cam (right).